\documentclass[pra,twocolumn,floatfix]{revtex4-1}
\usepackage{amsmath}
\usepackage{amsfonts}
\usepackage{graphicx}
\usepackage{color}
\usepackage{bm}
\usepackage{hyperref}

\begin{document}

\title{Environmental Coulomb blockade of topological superconductor-normal metal junctions}
\author{Konrad W\"olms and Karsten Flensberg}
\affiliation{Center for Quantum Devices, Niels Bohr Institute, University of Copenhagen, Juliane Maries Vej 30, 2100 Copenhagen, Denmark}

\date{\today }

\begin{abstract}
We study charge transport of a topological superconductor connected to different electromagnetic
environments using a low-energy description where only the Majorana bound states in the
superconductor are included. Extending earlier findings
who found a crossover between perfect Andreev reflection with conductance $2e^2/h$ to a regime
with blocked transport when the resistance of the environment is larger than $2e^2/h$, we consider
Majorana bound states coupled to metallic dots. In particular, we study two topological superconducting leads connected
by a metallic quantum dot in both the weak tunneling and strong tunneling regimes. For weak
tunneling, we project onto the most relevant charge states. For strong tunneling, we start from the Andreev
fixed point and integrate out charge fluctuations which gives an effective low-energy model for the non-perturbative
gate-voltage modulated cotunneling current. In both
regimes and in contrast to cotunneling with normal leads, the conductance is temperature
independent because of the resonant Andreev reflections, which are included to
all orders.
\end{abstract}

\maketitle

\affiliation{Center for Quantum Devices, Niels Bohr Institute, University of Copenhagen, 2100 Copenhagen, Denmark}

\section{Introduction}

There is currently a large attention towards topological superconductors, which
are expected to host zero-energy states with interesting properties. These
states, known as Majorana bound states (MBSs), constitute half fermions in the
sense that two MBSs are needed to define a single fermionic level, where the
degree of freedom of this fermion corresponds to the parity of the fermion
number. Thus, if the MBSs are well separated in space the parity degree of
freedom is a topologically protected quantity, which can potentially be used for
quantum computation \cite{Nayak2008}.

Promising candidate systems are hybrid materials where s-wave superconductors
are used to proximitizes other materials with strong spin-orbit coupling
\cite{BeenakkerReview,Alicea2010,LeijnseReview}. Systems like this have already
been studied
in a number of experiments \cite{Mourik2012,Das2012,Deng2012,Churchill2013}.

An important diagnostic tools to verify the presence of MBSs is tunnel
spectroscopy for which the zero-energy bound states are predicted to give
resonant Andreev reflection and hence a low-temperature conductance of
$2e^{2}/h$ \cite{Sengupta2001,Law2009,Flensberg2010}. With interactions in the
normal leads the situation changes because the Andreev reflection may be
suppressed by repulsive interactions in the metal. This was studied by Fidkowski
\textit{et al.} \cite{Fidkowski2012}, who showed that when the normal metal is a
Luttinger liquid, the resonant Andreev reflection is replaced by a
non-conducting fixed point for interactions stronger than a critical value. In
particular, the prediction is that for a chiral Luttinger liquid the cross-over
happens for the Luttinger parameter $K=\frac{1}{2}$, with the insulating phase
occurring for strong interactions, $K<\frac{1}{2}.$ More details of the voltage
and temperature dependence were studied later \cite{Lutchyn2013}. A related
study of a point contact between a Luttinger liquid and a chiral Majorana mode
was done in Ref.~\onlinecite{Lee2014}.

Another interesting situation is when the Majorana bound state is tunnel coupled
to an interacting quantum dot
\cite{Leijnse2011,Golub2011,Lee2013,Cheng2014,Dong2014}. For such a setup the
Coulomb blockade results in sharper resonance structures in the weak-coupling
limit \cite{Leijnse2011}. This was shown to hold true also for
stronger coupling  by Cheng \textit{et al.} \cite{Cheng2014}, who showed that
the perfect Andreev reflection dominates over the Kondo effect in the strong-coupling
fixed point.
\begin{figure}[tbp]
\includegraphics[width=.45\textwidth]{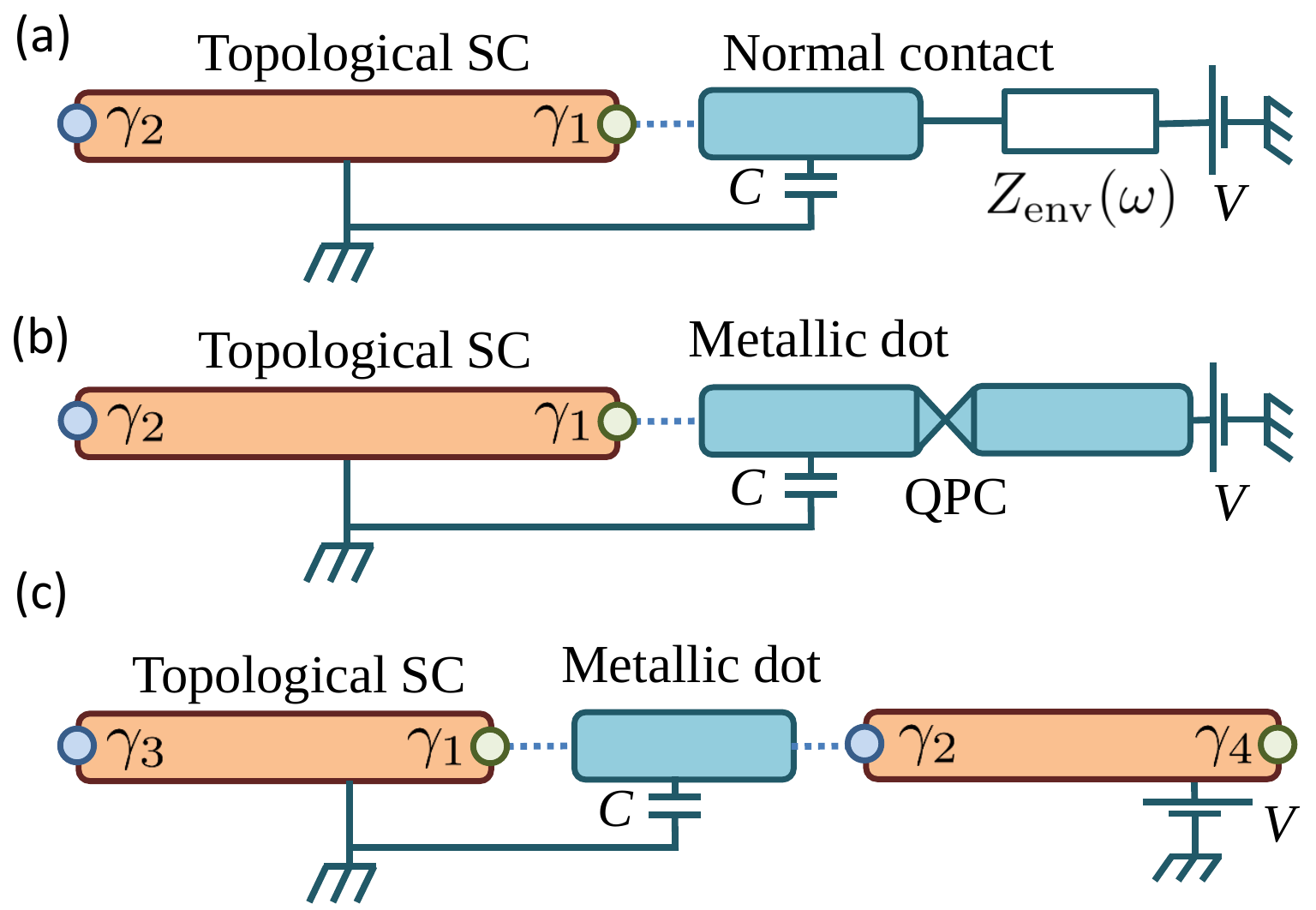}
\caption{Illustration of the three types of junctions considered here. In (a) there is a junction
between a topological superconductor (SC) and metallic lead with its electromagnetic environment
modeled by an impedance $Z_{\mathrm{env}}(\omega)$. In (b) the environment is replaced by a metallic island connected
to the drain by a quantum point contact. In (c) is a junction between two topological
superconductors, where the junction consists of a metallic Coulomb blockaded island.}
\label{fig:systems}
\end{figure}

In this paper, we analyze the transport properties through MBSs via different
types of environments, namely (a) a general linear electromagnetic environment,
(b) a metallic quantum dot with a quantum point contact drain, and (c) a
metallic dot with a second topological superconductor as a drain. See Fig.~\ref{fig:systems} for
illustrations. The situation in (a) was shown by Liu \cite{Liu2013} to exhibit a crossover from
conducting to non-conducting when the environment impedance is larger than the
resistance of the resonant Andreev conductance $Z_{\mathrm{env}}>h/2e^{2}$, and the
same results follows in the situation (b) when the quantum point contact has less than two open channels.
The nature of this transition is completely analogous to the Luttinger-liquid physics
studied by Fidkowski \textit{et al.} \cite{Fidkowski2012}.

The above transitions occur when the isolated MBS-metal junction resistance and the environment
resistance are equal. This motivates the study of situation (c) in Fig.~1 where two
MBS-metal junctions are connected via a Coulomb-blocked metallic island. Because
both junctions have resistance $h/2e^2$, the combined resistance is locked to
the transition value and the conductance is temperature independent, similar to a
non-interacting scattering problem. In fact, after integration over charge
fluctuations the resulting low-energy model can be
solved by mapping to a simple fermionic scattering problem by refermionization.

The paper is organized as follows. In Sec.~\ref{sec:bosonization}, we introduce
the key elements of the models, namely a single MBS coupled to a fermion bath
and the bosonic electromagnetic environment, and show how these elements are
bosonized. In Sec.~\ref{sec:environ}, the effective low-energy actions is
derived and used to identify the crossover impedance. The tunneling limit is
treated in detail in Sec.~\ref{sec:lineshape}. The case of a quantum point
contact is included as special case in Sec.~\ref{sec:qpc}. Next, the Coulomb
blockade of a metallic island coupled to two MBSs is analyzed in
Sec.~\ref{sec:CB} using the  dual bosonized representation. Finally, conclusions
are given in Sec.~\ref{sec:concl}

\section{Bosonization of the MBS-normal metal-environment contact}
\label{sec:bosonization}

We start with the Hamiltonian of a MBS tunnel coupled to a normal-metal lead,
with the inclusion of an displacement of the environment charge which
accompanies the tunneling event
\begin{equation}\label{eqn:HamiltonainElectronsEnvironmentAndTunneling}
    H=H_{\mathrm{el}}+H_{\mathrm{env}}+H_{\mathrm{T},\mathrm{env}},
\end{equation}%
where
\begin{align}
H_{\mathrm{el}}=&\sum_{k}\xi_k c_{k}^\dagger c_{k}^{{}},\\
H_{\mathrm{T},\mathrm{env}}=&t\sum_{k}\left( c_{k}^{{}}e^{i\alpha}-c_{k}^{\dagger}e^{-i\alpha}\right)\gamma,
\end{align}
and where $k$ are the quantum numbers for the normal-metal eigenstates that
couple to the MBS, $\gamma$, and $H_{\mathrm{env}}$ is the Hamiltonian of the
linear environment. The operator $\alpha$ is the canonically conjugate operator
to the environment charge $Q$, i.e. $\left[ \alpha,Q \right]=i$.
Therefore $e^{\pm i\alpha}$ creates/annihilates a charge in the environment
respectively. The current operator is given by
\begin{equation}
I=iet\sum_{k}\left( c_{k}^{{}}e^{i\alpha}+c_{k}^{\dagger}e^{-i\alpha}\right) \gamma.
\end{equation}
Because the MBS couples to a particular spin projection of the lead
\cite{Flensberg2010}, we only need to include the corresponding spin direction and
we can effectively use a spinless model.

Moreover, we take the tunneling matrix element $t$ to be independent of $k$
(s-wave scattering) and, furthermore, assume that the density of states is
constant: $\rho
(\xi)=\sum_{k}\delta \left( \xi -\xi _{k}\right) =\rho_{0}$. These
simplifications allows us to parameterize the dispersion relation as
$\xi_{k}=v_{\mathrm{F}}^{{}}\left(k-k_{\mathrm{F}}^{{}}\right) $ and map
$H_{\mathrm{el}}$ to a 1D chiral electron gas
\begin{equation}
    H_{\mathrm{el}}=v_{\mathrm{F}}\int_{-\infty }^{\infty }\mathrm{d}x~\psi ^{\dagger }(x)(-i\partial
    _{x}-k_{\mathrm{F}})\psi (x),
\end{equation}%
with
\begin{equation}
\psi (x)=\frac{1}{\sqrt{\mathcal{L}}}\sum_{k}e^{ikx}c_{k}^{{}},\quad
c_{k}^{{}}=\frac{1}{\sqrt{\mathcal{L}}}\int\! \mathrm{d}x~e^{-ikx}\psi (x)
\end{equation}%
and $\mathcal{L=}2\pi \rho _{0}v_{\mathrm{F}}^{{}}$ playing the role of a normalization length. In this
language, the tunneling Hamiltonian becomes%
\begin{equation}
    H_{\mathrm{T}}=\lambda \left( \psi (0)e^{i\alpha}-\psi ^{\dagger }(0)e^{-i\alpha}\right) \gamma ,
\end{equation}%
where $\lambda =\sqrt{2\pi \rho _{0}v_{\mathrm{F}}^{{}}t^{2}}.$

This 1D Hamiltonian can now be bosonized in the standard way, for example by folding into a
half-infinite wire, such that $\psi (x>0)=\psi _{\mathrm{R}}(x)$ and $\psi
(x<0)=\psi _{\mathrm{L}}(x).$ Following
Fidkowski \textit{et al.}, we integrate out all $x\neq 0$ degrees of freedom, which then gives the action%
\begin{eqnarray}
    S &=&S_0[\Theta]+S_{\mathrm{T}}[{\Theta},\alpha]+S_\mathrm{env}[\alpha], \\
S_0[\Theta] &=&\frac{1}{2\pi \beta }\sum_{i\omega _{n}}|\omega _{n}||\Theta
(i\omega _{n})|^{2}, \\
S_{\mathrm{T}}[{\Theta }]&=&2\lambda ^{\prime }\int_{0}^{\beta }\mathrm{d}\tau ~\sigma _{x}\cos\left(
\Theta (\tau )+\alpha (\tau )\right),\label{ST}
\end{eqnarray}%
where $\Theta (\tau )$ relates to current at $x=0$~in the half-wire language
(Fidkowski \textit{et al.} \cite{Fidkowski2012}) or to $\Theta
=\pi \int_{-\infty }^{\infty}\mathrm{d}x~\mathrm{sign}(x)\rho (x)$ in the 1D chiral language. Furthermore,
the Pauli matrix $\sigma _{x}$ refers to the two-level system defined in a complex-fermion basis
given by $\gamma $ and another MBS in the topological SC (for example $\gamma _{2}$ in the Fig.
1(a)). Also a short scale cut-off parameter $a $ has been included, so that $\lambda
^{\prime }=\lambda /\sqrt{2\pi a}$. Similarly, the current operator in its
bosonized form is given by
\begin{equation}
I(\tau )=2\lambda ^{\prime }~\sigma _{x}\sin \left(\Theta (\tau )
+\alpha(\tau)\right).
\end{equation}

\section{MBS coupled to an electromagnetic environment}
\label{sec:environ}

In this section, we study a MBS coupled to two different environments: firstly a general linear
impedance and then secondly a quantum point contact (QPC).

\subsection{MBS coupled to a linear environment}

The environment action describes the dynamics of the field $\alpha $ and thus after the integration over all other environment degrees of freedom, it
can be expressed as
\begin{equation}
S_{\mathrm{env}}[\alpha]=\frac{1}{2\beta }\sum_{i\omega _{n}}G_{\alpha \alpha
}^{-1}(i\omega _{n})|\alpha (i\omega _{n})|^{2},
\end{equation}%
where $\alpha$'s Green's function is defined as $G_{\alpha \alpha
}(\tau)=-\left\langle T_{\tau }\alpha (\tau )\alpha (0)\right\rangle $. The
$\alpha $ correlation now follows from classical circuit theory, analogous to
the environmental-Coulomb-blockade theory \cite{Girvin1990,
Devoret1990,IngoldNazarov}. In frequency domain we have
\begin{equation}
G_{\alpha \alpha }(i\omega _{n})=\frac{e^{2}}{|\omega _{n}|^{2}}%
G_{VV}(i\omega _{n})=\frac{e^{2}|Z_{\mathrm{t}}(i\omega _{n})|^{2}}{|\omega _{n}|^{2}}%
G_{II}(i\omega _{n}),
\end{equation}%
where $G_{VV}$ and $G_{II}$ are the voltage-voltage and current-current
correlation function,
respectively, which by the Kubo formula are related to the total impedance of the network as
$G_{II}=|\omega _{n}|Z_{\mathrm{t}}^{-1}$. We thus get an expression for $G_{\alpha \alpha }$ in terms of the
total environment impedance
\begin{equation}
    G_{\alpha \alpha }(i\omega _{n})=\frac{e^{2}Z_{\mathrm{t}}^{\ast }(i\omega _{n})}{%
|\omega _{n}|}.
\end{equation}%
The network consists of the junction capacitor and the environment impedance in parallel (see Fig.
1(a)) and therefore we obtain
\begin{equation}
G_{\alpha \alpha }(i\omega _{n})=\frac{e^{2}}{|\omega _{n}|\left( |\omega
_{n}|C+\left( Z_{\mathrm{env}}^{\ast }\right) ^{-1}\right) },
\end{equation}%
where $Z_{\mathrm{env}}^{{}}$ is the impedance of the network connecting the junction to the voltage source.

Now, from the action (\ref{ST}) we see that the field $\Theta -\alpha $ does not
couple to the junction and we can integrate it out. This leaves us with an
effective action for $\Theta _{+}=\Theta +\alpha$ where the quadratic part
becomes
\begin{equation}
S_{0}[\Theta _{+}]=\frac{1}{2\pi \beta }\sum_{i\omega _{n}}|\omega
_{n}|G_{+}^{-1}(i\omega_n)|\Theta _{+}(i\omega _{n})|^{2},
\end{equation}%
with
\begin{equation}
G_{+}(i\omega _{n})=1+\frac{|\omega_n|}{\pi }G_{\alpha \alpha}(i\omega _{n}),
\end{equation}%
and the non-quadratic part is given by Eq. (\ref{ST}).

Let us now study the case where the environment impedance can be approximated by a constant real
impedance. For small frequencies the Green's function $G_{+}(i\omega _{n})$ reduces to
\begin{equation}
G_{+}(i\omega _{n})\approx 1+\gamma ,
\end{equation}%
where we have defined $\gamma =2e^{2}Z_{\mathrm{env}}/h.$ The environment model is thus identical to
the situation of a MBS coupled to a Luttinger liquid with the replacement $K\rightarrow (1+\gamma
)^{-1}$. \emph{Thus, when the environment impedance exceeds }$h/2e^{2}$\emph{, corresponding to
}$\gamma >1$\emph{\ and }$K<1/2,$\emph{\ there is a transition from resonant Andreev reflection to
insulator behavior at low temperature and low voltage.}

\subsection{Line shape of Majorana environmental Coulomb blockade close to the non-conducting
fixed point}
\label{sec:lineshape}

It is also interesting to discuss the finite-voltage lineshape of the peak in
differential conductance, because this is most often the experimental signature
used to conclude on the presence of MBSs in the superconductor. The tunneling
current is calculated to second order in $t$ by standard perturbation theory:
\begin{equation}\label{I}
    I= -ie\int_{-\infty}^0 \mathrm{d}t\,\langle\{I(0),H_{\mathrm{T},\mathrm{env}}(t)\}\rangle.
\end{equation}
After some algebra, we then obtain
\begin{equation}\label{Ifinal}
    I= \frac{e\Gamma}{h}\int_{-\infty}^\infty \mathrm{d}\omega\left[f(\omega-eV)-f(\omega+eV)\right]P(\omega),
\end{equation}
where $f(\omega)=1/(\exp(\beta\omega)+1)$ is the Fermi-Dirac distribution function where we sat the
reference chemical potential of the superconductor to zero, $\Gamma=2\pi e^2t^2\rho$ (with $\rho$
being the density of states in the normal-metal contact) and where
\begin{equation}\label{Pdef}
    P(\omega)=\int_{-\infty}^\infty  \mathrm{d}t\,e^{i\omega t}\left\langle
    e^{i\alpha(t)}e^{i\alpha(0)}\right\rangle,
\end{equation}
is the usual ``$P(E)$-function'' \cite{Devoret1990,Girvin1990}, which obeys
$P(\omega)=P(-\omega)e^{\beta\omega}$ and describes the response of the environment. Without
coupling to the environment, we have $P(\omega)=2\pi\delta(\omega)$ and hence
$\mathrm{d}I/\mathrm{d}V=4\pi\Gamma
(2e^2/h) (-f'(eV))$. With an environment the differential conductance is
\begin{equation}\label{eqn:dIdV}
    \frac{\mathrm{d}I}{\mathrm{d}V}=\frac{e^2}{h} \Gamma\int_{-\infty}^\infty
    \mathrm{d}\omega\left(-\frac{\mathrm{d}f(\omega)}{\mathrm{d}\omega}\right)
 (P(\omega+eV)+P(\omega-eV)),
\end{equation}
which at zero temperature reduces to
\begin{equation}\label{eqn:dIdVZeroTemperature}
    \left. \frac{\mathrm{d}I}{\mathrm{d}V}\right|_{T=0}=\frac{2e^2}{h} \Gamma (P(eV)+P(-eV)).
\end{equation}
Interestingly, the normal-metal-MBS junction thus measures the $P(\omega)$, in contrast to the usual
metallic junction where the differential conductance is given by the integral of $P(\omega)$
\cite{Girvin1990}. At zero temperature and low energies, the environment function goes as
$P(\omega)\propto \omega^\gamma$, which confirms the conclusion from the previous section that for
$\gamma>1$, the zero-bias conductance goes to zero at low temperatures.

\subsection{Metallic dot with a quantum point contact drain}
\label{sec:qpc}

Next, we consider a special kind of environment, namely a metallic dot with connections to a MBS and
quantum point contact with $N$ open channels to the drain electrode, see Fig. 1(b). The environment
action includes the charging energy of the dot, the coupling to the phase field $\alpha$, and the
bosonized open channels of the quantum point contact. It reads%
\begin{eqnarray}
    S_{\mathrm{env}} &=&\int_{0}^{\beta }\mathrm{d}\tau ~\left( i\dot{\alpha}Q+\frac{e^{2}%
}{2C}\left( Q+\frac{1}{\pi }\sum_{i=1}^{N}\phi _{i}-N_{\mathrm{g}}\right) ^{2}\right)
\notag \\
&&+\frac{1}{\pi \beta }\sum_{i,n}|\omega_n||\phi _{i}|^{2},
\end{eqnarray}%
where $Q$ is the conjugate field to $\alpha $, counting the number of charges passing through the Majorana-bound-state junction, and $\phi _{i}$
is the boson field describing the charge that passed from lead to dot through mode $i$ in the QPC \cite{Flensberg1993,Matveev1995}, and $N_{\mathrm{g}}$ is
controlled by a gate voltage. After integrating out all $\phi _{i}s $ as well as the field $Q$ (and remove $N_{\mathrm{g}}$ by a gauge
transformation) the environment action describing the field $\alpha $ becomes identical to the electromagnetic model above with the environment
impedance given by $Z_{\mathrm{env}}=h/Ne^{2}$ or in terms of the parameter $\gamma $ defined above, we get an effective $\gamma $ for the QPC setup
given by%
\begin{equation}
    \gamma _{\mathrm{QPC}}^{{}}=2/N.
\end{equation}%
\emph{This shows that the there is a crossover between full normal reflection and full Andreev reflection at }$N=2.$

\section{Two MBS coupled via a metallic dot}
\label{sec:CB}

Finally, we consider the device setup in Fig. 1(c) consisting of a metallic Coulomb-blocked island
coupled to two MBSs. It is an inverse version of the system studied by Fu
\cite{Fu2010} and later by Hutzen \textit{et al.} \cite{Hutzen2012,Zazunov2014}, namely a topological
superconductor Coulomb island with the two MBSs coupled to normal leads. The setup allows for a non-perturbative solution of the conductance and therefore gives an interesting way to
investigate the combination of MBSs and Coulomb interactions. The Hamiltonian of the metallic dot coupled to two MBS is
\begin{equation}\label{eqn:HMBSdot}
    H=\sum_{r=\mathrm{L},\mathrm{R}}
    H_{\mathrm{el},r}+H_{\mathrm{T},r}+H_{\mathrm{C}},
\end{equation}%
with
\begin{align}
    H_{\mathrm{el},r}&=\sum_{k}\xi_k c_{k,r}^\dagger c_{k,r}^{{}}\\
    H_{\mathrm{T},r}&= t_r\sum_{k}\left( c_{k,r}^{{}}-c_{k,r}^{\dagger}\right) \gamma_r,\\
    H_{\mathrm{C}}&=\,E_{\mathrm{C}}(N_{\mathrm{L}}+N_{\mathrm{R}}-N_{\mathrm{g}})^2,
\end{align}
where $r=\mathrm{L},\mathrm{R}$ labels the two ends of the quantum dot, which
have independent quantum numbers labeled by $k$, and the MBSs are denoted by
$\gamma_{\mathrm{L},\mathrm{R}}$. The current operators at the two contacts are
\begin{equation}
    I_r=iet_r\sum_{k}\left( c_{k,r}^{{}}+c_{k,r}^{\dagger}\right)
    \gamma_{r}.
\end{equation}

For weak tunneling and away from $N_{\mathrm{g}}$ being a half-integer, the dot
has a well defined charge and current is therefore blocked as in the usual
Coulomb blockade. In this situation, current is carried by cotunneling, which is
via virtual states with one additional charge. Second-order perturbation theory
then gives the following result for the conductance at $T\ll E_{\mathrm{C}}$
\begin{equation}\label{Gcotun}
    G_\mathrm{cotun}=\frac{e^{2}}{h}\Gamma_\mathrm{L}\Gamma_\mathrm{R}
\left(\frac{1}{E_{\mathrm{C}}^+}+\frac{1}{E_{\mathrm{C}}^-}\right)^2,
\end{equation}
where $\Gamma_r=2\pi v_{\mathrm{F}} |t_r|^2$ and
$E_{\mathrm{C}}^\pm=E^{\phantom{\pm}}_{\mathrm{C}}(1\mp2 N_{\mathrm{g}})$, with
$N_{\mathrm{g}}\in [-1/2,1/2]$. We note that the cotunneling current is
independent of temperature at low temperatures. The cotunneling conductance
diverges near the charge degeneracy points $N_{\mathrm{g}}=\pm 1/2$. To
understand the behavior near these points, we follow the discussion by Fu
\cite{Fu2010}. For large charging energy $E_{\mathrm{C}}\gg\Gamma_r$, we can restrict the
analysis to two charge states, for example near $N_{\mathrm{g}}=1/2$, where
$N_{\mathrm{L}}+N_{\mathrm{R}}$ is either 0 or 1. The charge on the dot can then be represented by a
fermionic degree of freedom: $N=N_{\mathrm{L}}+N_{\mathrm{R}}=f^\dagger f$, such that
\begin{equation}\label{fdef}
    (N_{\mathrm{L}}+N_{\mathrm{R}}-N_{\mathrm{g}})^2=f^\dagger f(1-2N_{\mathrm{g}})-N_{\mathrm{g}}^2.
\end{equation}
We can also express the tunneling Hamiltonian in terms of the fermion $f$. This
is done by observing that $[c_{k,r}^\dagger\gamma_r,N]=c_{k,r}^\dagger\gamma_r$, such
that the same dynamics is obtained if we replace $c_{k,r}^\dagger\gamma_r$ by
$c_{k,r}^\dagger f$ and treat $c_{k,r}^{\dagger}$ and $f$ as independent
fermions, because $[c_{k,r}^\dagger f,f^\dagger
f]=c_{k,r}^\dagger f$. The Hamiltonian that governs the system projected onto
the two charge states, 0 and 1, can thus be written as
\begin{equation}\label{eqn:HMBSdotProjected}
H_{\mathrm{P}}=\sum_{r=\mathrm{L},\mathrm{R}}
\left[H_{\mathrm{el},r}+H_{P,\mathrm{T},r}\right]+E_{\mathrm{C}} f^\dagger f(1-2N_{\mathrm{g}}),
\end{equation}%
with
\begin{equation}
    H_{\mathrm{P},\mathrm{T},r}= t_r\sum_{k}\left( f^\dagger c_{k,r}^{{}}+c_{k,r}^{\dagger}f\right).
\end{equation}
Likewise, the current operators becomes
\begin{equation}
I_r=-iet_r\sum_{k}\left(f^\dagger c_{k,r}^{{}}-c_{k,r}^{\dagger}f\right).
\end{equation}
The projected Hamiltonian describes a resonant level, $f$, coupled to electron
reservoirs, $r=\mathrm{L}$ and $r=\mathrm{R}$. The conductance is then simply given by the conductance of a resonant level
\begin{equation}\label{Gp}
    G_\mathrm{P}=\frac{e^{2}}{h}\frac{4\Gamma_\mathrm{L}\Gamma_\mathrm{R}}{4E_{\mathrm{C}}^2
    (1-2N_{\mathrm{g}})^2+\Gamma^2+2\Gamma_{\mathrm{L}}\Gamma_{\mathrm{R}}},
\end{equation}
where $\Gamma^2=\Gamma_\mathrm{L}^2+\Gamma_\mathrm{R}^2$. Eq.~\eqref{Gp}
agrees with the perturbative cotunneling result in Eq.~\eqref{Gcotun} if only
cotunneling via the $N=1$ (and not $N=-1$) charge is included, i.e. keeping only
the $E_{\mathrm{C}}^+$ term in \eqref{Gcotun}. Parameterizing the
$\mathrm{L}/\mathrm{R}$ asymmetry as
$\sin\theta_{\mathrm{P}}=\Gamma_{\mathrm{L}}/\Gamma$ and
$\cos\theta_{\mathrm{P}}=\Gamma_{\mathrm{R}}/\Gamma$, we can write
the conductance in the charge-projected basis as
\begin{equation}\label{Gptheta}
    G_{\mathrm{P}}=\frac{e^{2}}{h}\frac{2\sin
        2\theta_{\mathrm{P}}}{1+ \sin2\theta_P+(2E_{\mathrm{C}}(1 -2N_{\mathrm{g}})/\Gamma)^2}.
\end{equation}

The non-perturbative result in \eqref{Gp} was derived under the assumption that
the charge fluctuations are suppressed to the two charge states closest in
energy, or in other words, $E_{\mathrm{C}}\gg\Gamma_r$. In order to investigate the
opposite limit where contacts have large transparency, we describe the Majorana
junction in terms of a dual representation, which is an expansion around the
perfect Andreev
fixed point \cite{Fidkowski2012}. In the dual representation, the action of the two Majorana junctions reads\cite{Fidkowski2012}
\begin{align}\notag
    S_{\mathrm{dual}}^{{}}[\Phi_{\mathrm{L}},\Phi_{\mathrm{R}}]=&\frac{1}{2\pi \beta }\sum_{\substack{i\omega
        _{n}\\r=\mathrm{L},\mathrm{R}}}|\omega
_{n}||\Phi _{r}(i\omega _{n})|^{2}\\
&+\sum_{r=\mathrm{L},\mathrm{R}}2\lambda_{r,\mathrm{bs}}^{{}}\int_{0}^{\beta }\mathrm{d}\tau ~\cos
2\Phi _{r}(\tau )+S_{\mathrm{C}},
\label{Sdualdot}
\end{align}
with
\begin{equation}
    S_{\mathrm{C}}[\Phi_{\mathrm{L}},\Phi_{\mathrm{R}}]=E_{\mathrm{C}}\int_{0}^{\beta
    }\!\mathrm{d}\tau\, \left(
    \frac{\Phi _{\mathrm{L}}}{\pi }+\frac{%
        \Phi _{\mathrm{R}}}{\pi }-N_{\mathrm{g}}\right) ^{2},
\end{equation}%
where the fields $\Phi_r$ give the charge that passes through junction $r$. It
is natural to introduce a total-charge field $\Phi_+=\Phi_{\mathrm{R}}+\Phi_{\mathrm{L}}$ and a
difference field $\Phi_-=\Phi_{\mathrm{R}}-\Phi_{\mathrm{L}}$, which is related to current via
$I=-i\partial_{\tau }\Phi_-/2\pi$. For $E_{\mathrm{C}}\gg \lambda_r$ and at low energies
$|\omega _{n}|\ll
E_{\mathrm{C}}$ the mode $\Phi_{+}$ gets pinned at
$\pi N_{\mathrm{g}}$. We can therefore integrate out the total-charge field by
replacing the cosine terms in the action \eqref{Sdualdot} by their averages over
$\Phi_+$. This is valid, when $E_{\mathrm{C}},\Lambda\gg \lambda_r$, where $\Lambda$ is the
high-energy cut-off parameter. This leaves the following effective low-energy
action for the difference field
\begin{align}\notag
    S[{\Phi _{-}}]=&\frac{1}{4\pi \beta }\sum_{i\omega _{n}}|\omega
_{n}||\Phi _{-}(i\omega _{n})|^{2}\\
&+\int_{0}^{\beta }\mathrm{d}\tau ~\left( \tilde{\lambda}_{\mathrm{bs}}^{{}}e^{i\Phi _{-}(\tau )}+\mathrm{c.c.}\right)\label{SPhim}
\end{align}%
where%
\begin{equation}
    \tilde{\lambda}_{\mathrm{bs}}^{{}}=\left( \lambda_{\mathrm{L},\mathrm{bs}}^{{}}e^{i\pi N_{\mathrm{g}}}+
    \lambda_{\mathrm{R},\mathrm{bs}}^{{}}e^{-i\pi N_{\mathrm{g}}}\right) \left\langle e^{i\Phi_{+}}\right\rangle_{0}.
\end{equation}
The expectation value  $\left\langle e^{i\Phi
_{+}}\right\rangle _{0}$ appears after integration over $\Phi_+$ and it becomes
\begin{equation}
\left\langle e^{i\Phi _{+}}\right\rangle _{0}=\exp \left( -\frac{\pi ^{2}}{%
\beta }\sum_{i\omega _{n}}\frac{1}{\pi |\omega _{n}|+4E_{\mathrm{C}}}\right) \approx
\frac{4E_{\mathrm{C}}}{\pi \Lambda },
\end{equation}%
where $\Lambda=v_{\mathrm{F}}/a$ is the cut-off energy. The model \eqref{SPhim}
is now solvable because it maps to a single non-interacting QPC
\cite{Flensberg1993}. The spirit of this mapping is equivalent to the solution
by Furusaki and Matveev \cite{FurusakiMatveev1995}, but while they mapped two
QPC coupled to a metallic dot to a Majorana junction, we here map two Majorana
junctions connected in series via a metallic dot to a single QPC. The
backscattering matrix element in the QPC model translates to
$V_{\mathrm{bs}}=2\pi a\tilde{\lambda}_{\mathrm{bs}}$. Moreover, the current
operator is $I=-i\partial
_{\tau }\Phi_-/2\pi$ for both the original model and for the QPC model.
Therefore, we can directly obtain the conductance for the dual model from the
QPC result as
\begin{equation}
    G_\mathrm{dual}=\frac{e^{2}}{h}\frac{1}{1+|V_{\mathrm{bs}}/2v_{\mathrm{F}} |^{2}},
\end{equation}%
and hence
\begin{align}\label{Gdual}
    G_\mathrm{dual}=\frac{e^{2}}{h}\frac{1}{1+(8E_{\mathrm{C}}
    /\Lambda^2)^2(\lambda_{\mathrm{bs}}^2+2\lambda_{\mathrm{L},\mathrm{bs}}^{{}}\lambda_{\mathrm{R},\mathrm{bs}}^{{}}\cos2\pi N_{\mathrm{g}})},
\end{align}
where
$\lambda_{\mathrm{bs}}^2=\lambda_{\mathrm{L},\mathrm{bs}}^2+\lambda_{\mathrm{R},\mathrm{bs}}^2$.
We also define an device asymmetry angle for the weak backscattering limit:
$\sin\theta_\mathrm{dual}=\lambda_{\mathrm{R},\mathrm{bs}}/\lambda_{\mathrm{bs}}$ and
$\cos\theta_\mathrm{dual}=\lambda_{\mathrm{L},\mathrm{bs}}/\lambda_{\mathrm{bs}}$,
so that
\begin{align}\label{Gdualtheta}
    G_\mathrm{dual}=\frac{e^{2}}{h}\frac{1}{1+(8E_{\mathrm{C}} \lambda_{\mathrm{bs}}/\Lambda^2)^2(1+\sin2\theta_\mathrm{dual}\cos2\pi N_{\mathrm{g}})},
\end{align}

\begin{figure}[t]
\begin{center}
\vspace*{.5cm}
\includegraphics[width=.45\textwidth]{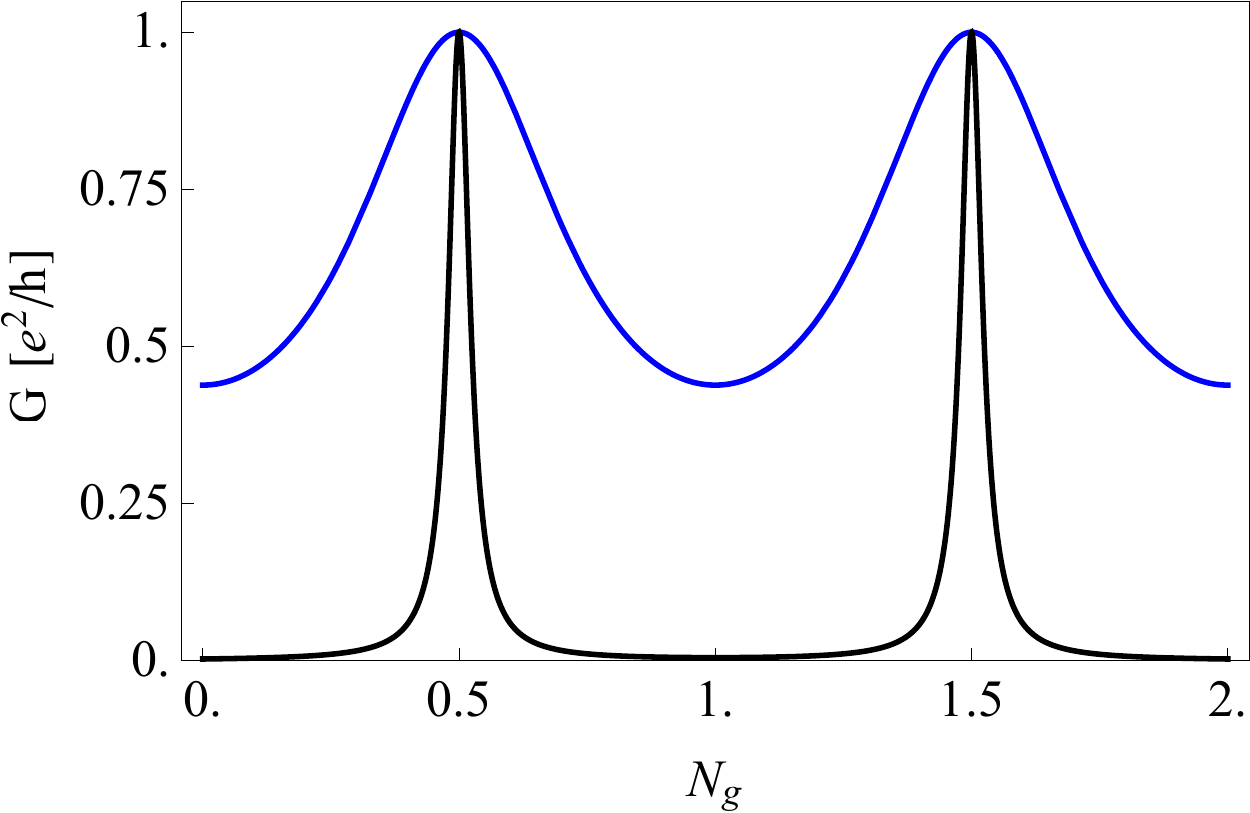}
\end{center}
\caption{The solutions for the conductance of a metallic island connected in the
    two MBS. The top curve is the result for the weak backscattering
    formulation (Eq.~\eqref{Gdualtheta}) for a symmetric device with
    $E_{\mathrm{C}}\lambda_\mathrm{bs}/\Lambda^2=0.1$, while the lower curve is the
    result for weak tunneling formulation (Eq.~\eqref{Gptheta}) with
    $\Gamma/E_{\mathrm{C}}=0.1$, also for a symmetric device. For the weak tunneling case
    we have added two curves, one centered at $N_\mathrm{g}=1/2$ and one
    centered $N_\mathrm{g}=3/2$, corresponding to projection to $N=0,1$ and
$N=1,2$, respectively.}
\label{fig:Gres}
\end{figure}

We see that in both the weak tunneling and in the weak backscattering
formulations the conductance peaks at the charge neutrality points and reaches
$e^2/h$ for a symmetrically coupled dot, agreeing with the expectation for the
series resistance of two resistors each having resistance $h/2e^2$. However, in
the general case they have different lineshapes. Fig.~\ref{fig:Gres} shows the
resulting conductance for the charge projected model and the dual model.

It is interesting to note that the conductance for the metallic dot coupled to
two MBSs does not depend on temperature, even though it is an inelastic process.
This is in contrast to the usual inelastic cotunneling in metallic dots, where
the conductance is proportional to temperature squared, even in the
strong-coupling limit \cite{FurusakiMatveev1995}.

\section{Summary and Conclusions}
\label{sec:concl}

Tunneling spectroscopy is an important diagnostic tool for determining the presence and properties of MBSs. In this paper, we have studied the influence of interactions with the surrounding circuit on the measured current-voltage characteristics. Confirming the results of Ref.~\onlinecite{Liu2013}, we obtain a transition from perfect Andreev reflection to an isolating state of the normal-metal-MBS junction which can be controlled by the resistance of the circuit. When the environment resistance exceeds $h/2e^2$, the linear conductance goes to zero at low temperatures. We studied two types of environments, namely one that can be described by a linear impedance and a quantum point contact, which allows for direct tuning of the transition.

Moreover, we introduced an interesting setup where two MBSs are coupled to a metallic island. This system has the property that the linear cotunneling current is temperature independent, which is very different from the usual cotunneling current through a metallic dot coupled to two metallic electrodes. In the latter case, the cotunneling current is proportional to temperature squared due to the phase space of final states with an electron-hole pair created at each contact. In contrast, with two MBS junctions connecting the dot, the integral over final states involving a single electron-hole pair also involves the divergent resonant Andreev reflections and, as a result, produces a temperature independent conductance. The metallic dot case is related to Coulomb blockade of a topological superconducting island considered by Fu \cite{Fu2010} and it can be solved in a similar way after projection onto two charge states. Furthermore, the temperature independent result turns out to be linked to a mapping to an effective low-energy model consisting of a 1D non-interacting fermion model with a single impurity. This mapping allowed us to derive a non-perturbative result for the low-temperature Coulomb blockade traces as a function of gate voltage. Finally, it is interesting to note that if the
metallic island in this setup  has more connecting MBSs , the system is an "inverse out" version of the topological Kondo effect system recently analyzed by several authors \citep{Beri2012,Beri2013,Altland2013}.

\begin{acknowledgments}
The research was supporting by the Danish National Research Foundation and by the Danish Council for
Independent Research \textbar{} Natural Sciences.
\end{acknowledgments}

%

\end{document}